\documentclass[11pt]{article}
\usepackage[margin=1in]{geometry}
\usepackage{amsmath}
\usepackage{mathptmx}
\usepackage{amsfonts}
\usepackage{float}
\usepackage{flafter}
\usepackage{subfigure}
\usepackage{varwidth}
\usepackage{graphicx}
\usepackage[T1]{fontenc}

\begin{document}
\title{Some New Approaches to MPI Implementations}
\author{Yuqing Xiong\\
(Computer Science Department, Shanghai Institute of Technology, Shanghai, China)\\
\texttt{yqxiong@sit.edu.cn}}
\date{}

\maketitle

\begin{abstract}
This paper provides some new approaches to MPI implementations to improve MPI performance. These approaches include dynamically composable libraries, reducing average layer numbers of MPI libraries, and a single entity of MPI-network, MPI-protocol, and MPI.
\end{abstract}

\section{Introduction}
MPI will be one of the most important components in exascale computing platforms in the future. The performance of MPI is very important to parallel applications. In this paper, we will present some new approaches to MPI implementations, including dynamically composable libraries, reducing average layer numbers of MPI libraries, and a single entity of MPI-network, MPI-protocol, and MPI, to improve MPI performance.

\section{Dynamically Composable Libraries}
\subsection{Basic Principle}
Many applications (such as the BLACS[1]) usually only invoke subsets of MPI function set, a lot of functions of MPI aren't invoked in their entire run period. Thus, MPI implementations can be designed to be dynamically composable libraries, that is, the functions invoked by applications form dynamically ``thin'' temporary MPI libraries only for the applications (in other word, all MPI functions invoked in a application form a temporary MPI library only for the application, i.e., one application corresponds to one library).

For example, assume that there is an application, and that the functions invoked in the application are MPI\_Init, MPI\_Comm\_size, MPI\_Comm\_rank, MPI\_Send, MPI\_Recv, and MPI\_Finalize. These six functions can form a dynamically composable MPI library only for the application. When the application finishes its execution, the library will be removed.

It should be believed that there would be a performance advantage to a dynamically composable MPI library since it contains only a subset of MPI function set. Let's look at an extreme example, assume that there two MPI libraries, $\mathcal{A}$ and $\mathcal{B}$. $\mathcal{A}$ consists of \{MPI\_Init, MPI\_Comm\_size, MPI\_Comm \_rank, MPI\_Send, MPI\_Recv, MPI\_Finalize\}, and $\mathcal{B}$ contains all MPI functions. Obviously, it is possible that we can implement $\mathcal{A}$ from the ground up and make its performance higher than $\mathcal{B}$. The goal of dynamically composable MPI libraries is exactly to produce the implementation of $\mathcal{A}$ automatically.

\subsection{About Implementation of the Approach}
The word ``dynamically'' in a dynamically composable library means for different applications or on demand at application execution time. So there are different implementations for it. We give an approximate approach to implementing a dynamically composable library for the previous implication. The approach is to built the library before an application execution. Before the application execution, the application code is scanned to record invoked MPI functions, which is similar to lexical analysis of compilers, then a dynamically composable library containing the invoked MPI functions is formed and installed, finally we can compile, link, and run the application based on the dynamically composable library.

Dynamically composable MPI libraries may be implemented like toy building blocks. All necessary basic blocks should be implemented in advance. Every dynamically composable MPI library can be temporarily built from these blocks before any application execution.

In order to do this, we divide the set of all MPI functions into \emph{n} subsets $F_1$, $F_2$, $\cdots$, $F_n$ according to functionalities (or other rules). These $F_i$ ($i=1,\cdots,n$) correspond to the basic blocks.

Assume that \emph{$\mathcal{F}$} is a set of MPI functions invoked by an application, and that \emph{m} is such a minimum number that \emph{$\mathcal{F}$} $\subseteq$ $F_{i_1}$ $\bigcup$ $F_{i_2}$ $\bigcup$$\cdots$$\bigcup$ $F_{i_m}$. A dynamically composable library for the application is $F_{i_1}$ $\bigcup$ $F_{i_2}$ $\bigcup$$\cdots$$\bigcup$ $F_{i_m}$.

\section{Reducing Average Layer Numbers of MPI Libraries}
MPI implementations are generally designed in term of multi-layer structures, see Figure 1-A. Obviously, the smaller is layer number of MPI library, the higher is its performance.

\begin{figure}[h]
\begin{center}
\includegraphics[scale=0.40]{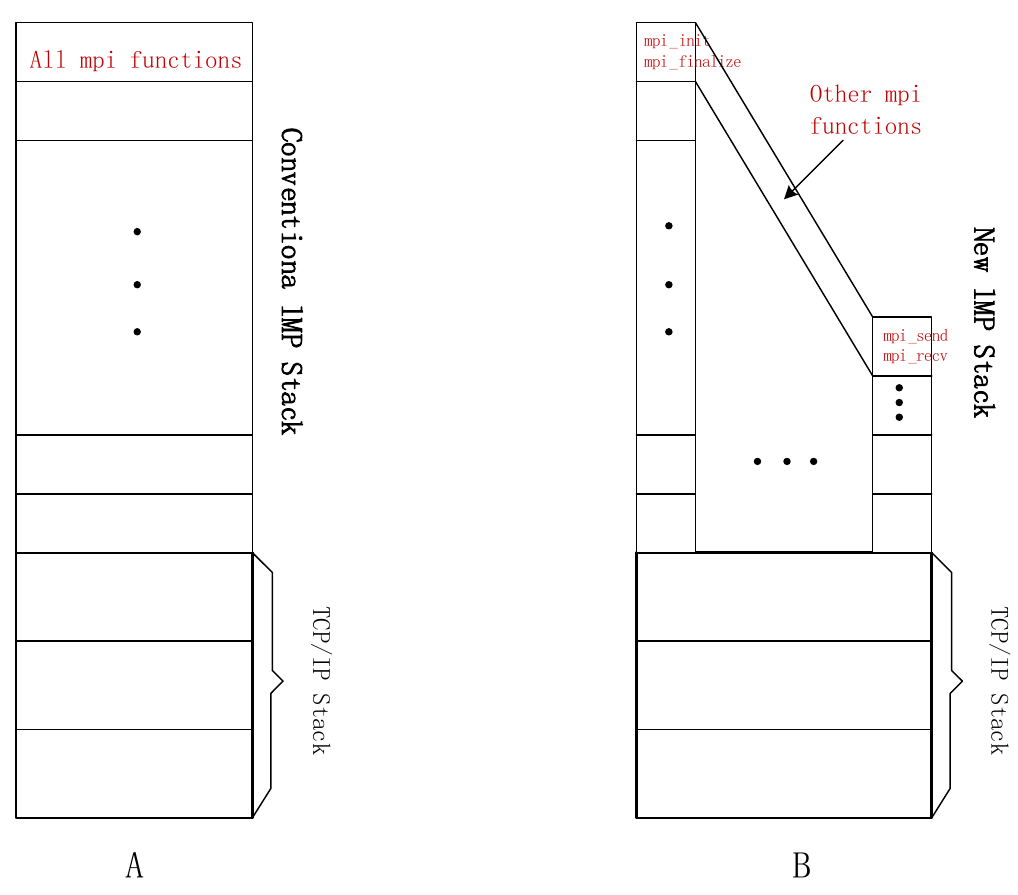}
\caption{MPI Stack}
\end{center}
\end{figure}

In order to reduce the layer number of MPI libraries, we can select some representative applications from key domains[2], and caculate global frequency of invocation of each MPI function in these applications statistically. The lower is the frequency of a function invocation, the larger is the layer number where the function stays at MPI stack, see Figure 1-B.

The above means that layer number of every MPI function in MPI stack may be different, which is much different from the conventional MPI stacks where all functions stay at the same layers.

For example, MPI\_INIT and MPI\_FINALIZA are invoked only once in an application, so the frequencies of their invocation are the lowest in MPI, while MPI\_SEND and MPI\_RECV may be two of the most invoked the functons. So MPI\_INIT and MPI\_FINALIZA should stay at the highest layer while MPI\_SEND and MPI\_RECV should stay at the lowest layer, see Figure 1-B, where we assume that MPI are based on TCP/IP. Thus, the average layer number of MPI implementation will be smaller in the new MPI stack architecture.

Therefore, the performance of applications run on the new MPI stack should be higher than on the conventional MPI stack.

\section{Single Entity of MPI, MPI-protocol, and MPI-network}
To cope with the challenges of exascale computing, some new operating systems, such as the Hobbes[3] and Argo[4], are designed from the ground up. So there are opportunities to give up TCP/IP and to design new communication protocol (such as Portal 4[5]) only for MPI, that is, the new communication protocol serves MPI whole-heartedly, and other communication services in exascale computing systems are carried out by other communication protocols (such as TCP/IP etc.). Therefore, we can regard the new protocol and MPI as a single entity, and call the new communication protocol MPI-protocol here.

\begin{figure}[h]
\begin{center}
\includegraphics[scale=0.46]{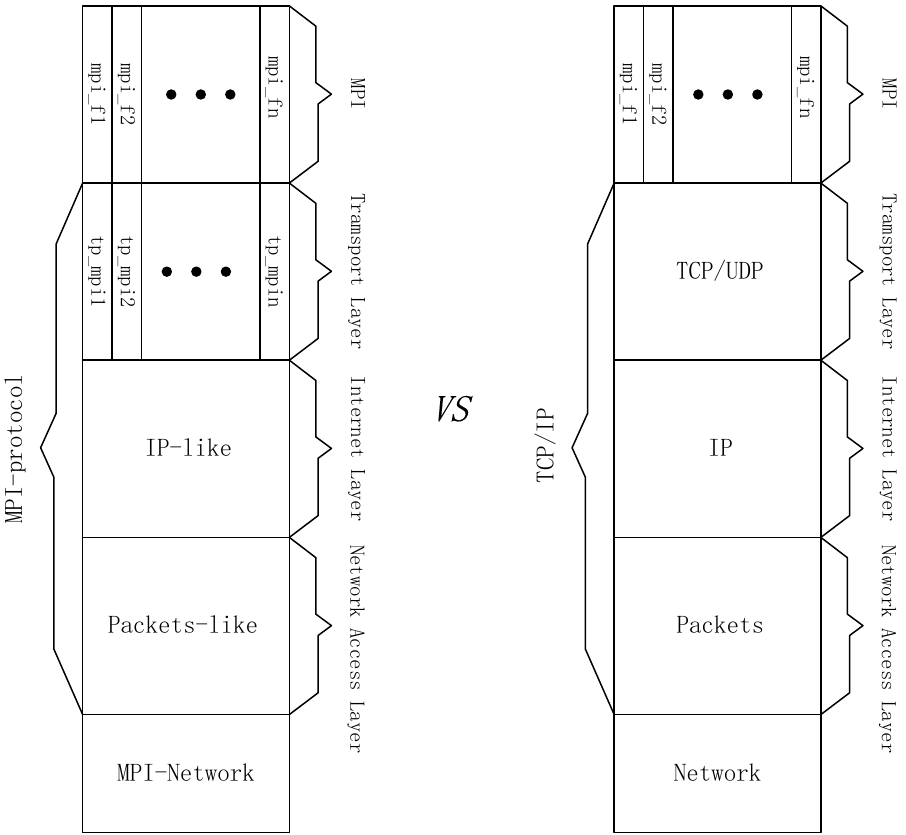}
\caption{MPI-network + MPI-protocol + MPI vs. Network + TCP/IP + MPI}
\end{center}
\end{figure}

MPI-protocol should be a set of communication protocols. In order to get high performance MPI, we can design a transport protocol for \textbf{every} MPI function which belongs to the MPI-protocol set, see Figure 2. Thus we have a loose space to design MPI-protocol and implement MPI. We can design and implement different communication protocols according to features and characteristics of MPI functions. For example, we can inject some important functionalities, such as fault tolerance and energy efficiency, into the protocols.

MPI implementation should include MPI-protocol stacks implementation ideally, that is, we should take MPI-protocol and MPI as a single entity to design and implement MPI.

In order to make performance of MPI-protocol and MPI higher, networks in HPC machines should be designed in speciality for MPI-protocol and MPI. We can call the networks MPI-networks, see Figure 2.

\section{Conclusions}
We summarize the above discussion as follows:
\begin{itemize}\itemsep=-2pt
\item{Dynamically composable MPI libraries are an approach to reducing the layer numbers of MPI libraries, and make MPI performance higher.}
\item{To reduce average layer numbers of MPI libraries to improve MPI performance, MPI functions can stay at different layers in MPI stack according to their invocation frequencies.}
\item{In order to get high performance MPI, we can design a communication protocol for each MPI function.}
\item{MPI-network, MPI-protocol, and MPI should be taken together as a single entity to design and implement in order to make the performance of MPI implementation higher.}
\end{itemize}
The three new approaches to MPI implementations do not conflict with each other, that is, they can be combined to applied to a single implementation of MPI. However, how to implement them will be a challenge.

\section*{Acknowledgment}

The author would like to thank Dr. Pavan Balaji and other researchers at Mathematics and Computer Science Division, Argonne National Laboratory. The author was invited by Dr. Pavan Balaji to visit Argonne in 2013, and some ideas of this paper were formed preliminarily at that time.

\vspace{6mm}
\noindent\rule[0.25\baselineskip]{\textwidth}{1pt}

\noindent BibTeX entry for citing the paper
\vspace{3mm}
\\
\texttt{
@inproceedings\{xiong2018,\\
    author=\{Yuqing Xiong\},\\
    title=\{Some New Approaches to MPI Implementations and a Possible Path to MPI Evolution\},\\
    booktitle=\{Proceedings of 11th International Congress on Image and Signal Processing, BioMedical Engineering and Informatics\},\\
    address=\{Beijing, China\}\\
    month=\{October\},\\
    year=\{2018\}\}
}
\end{document}